\documentclass{article}
\usepackage{epsfig}

\begin{document}

\title{Jet Formation from Rotating Magnetized Objects }

\author{G.S.Bisnovatyi-Kogan\thanks
{Space Research Institute, Profsoyuznaya str. 84/32, Moscow
117810, Russia, E-mail: gkogan@mx.iki.rssi.ru} \and
N.V.Ardeljan\thanks {Department of Computational Mathematics and
Cybernetics, Moscow State University, Vorobjevy Gory, Moscow B-234
119899, Russia, E-mail: ardel@redsun.cs.msu.su} ,\and
S.G.Moiseenko
\thanks {Space Research Institute, Profsoyuznaya str. 84/32,
Moscow 117810, Russia, E-mail:  moiseenko@mx.iki.rssi.ru}}
\date{}
\maketitle

\begin{abstract}
Jet formation is connected most probably with matter acceleration
from the vicinity of rotating magnetized bodies. It is usually
related to the mass outflows and ejection from accretion disks
around black holes. Problem of jet collimation is discussed.
Collapse of
a rotating magnetized body during star formation or supernovae
explosion may lead to a jet-like mass ejection for certain angular
velocity and magnetic field distributions at the beginning of the
collapse. Jet formation during magnetorotational
explosion is discussed basing on the numerical simulation of
collapse of magnetized bodied with quasi-dipole field.
\end{abstract}

%\maketitle

\section{Introduction}

Matter outflow is observed in most astrophysical objects in the
form of stellar wind, or collimated outflow (ejection) from young
stellar objects, AGN and quasars, microquasars (galactic X-ray
sources). Mechanisms of mass loss are connected with a radiative
and/or electromagnetic acceleration. The last one, which is
probably the best for producing collimated outflows, may work
quasi-stationary, or may be connected with explosive events.
Rotation is always present in compact objects, and formation of
collimated jets results from the action of magneto-rotational
phenomena.

\section{Problem of collimation}

During accretion and outflow of matter the relative dynamical
action of magnetic field increases for a given element of matter.
As was shown in \\ \cite{sch}, the conservation of a magnetic flux
during stationary accretion implies a dependence $B_r \sim
r^{-2}$. At constant mass flux $\dot M=4\pi \rho v_r r^2$ and
free-fall velocity $v_r \sim r^{-1/2}$, the density increases as
$\rho \sim r^{-3/2}$, and kinetic energy density $E_k \sim
r^{-5/2}$. The growth of the magnetic energy density is faster
$E_m \sim B_r^2 \sim r^{-4}$. After equipartition $E_k \sim E_M$
is reached, it cannot be violated during subsequent accretion.

 The outflow from rotating magnetized object contain
azimuthal component of the magnetic field, which energy density
decreases not faster than the specific kinetic energy of matter.
For the outflow with constant velocity $v_r$ in stationary
spherical outflow the density $\rho \sim r^{-2}$, $B_{\phi} \sim
r^{-1}$. In this conditions $E_k \sim E_m$, and the trajectories
of mass outflow become more tightly spiraled, and finally the flow
is becoming highly collimated with direction of the flow along the
rotational axis \cite{hn}, \cite{bog}.
We may say here about the universal magnetic
collimation in the outflows from rotating magnetized objects.

  The outbursts may be collimated at the very beginning. When jets
are separated from the object of its origin the problem appears of
preservation of the jet against its spherization during a motion
in a rarefied medium. One of the plausible mechanism of jet
preservation is a magnetic pinch collimation produced by an axial
electrical current, suggested in \cite{bkfk}.

\section{Jet formation by matter outflow from magnetized accretion
disk}

Accretion of matter with a large scale magnetic field into a black
hole leads to formation of an accretion disk with a strong
poloidal magnetic field. Models of non-rotating accretion disks,
supported by magnetic field had been constructed in
\cite{bkruz74}, \\ \cite{bkruz76}, see Figure 1.

\begin{figure}
\includegraphics[height=.5\textheight]{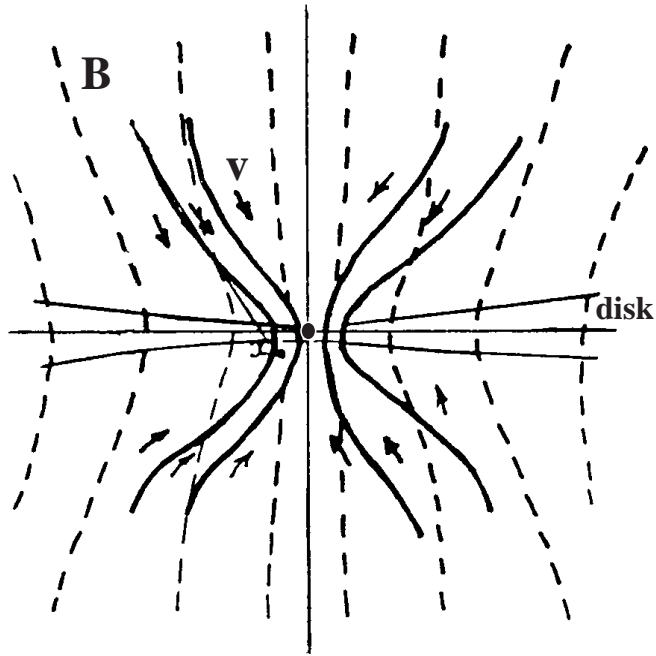}
\caption{Schematic picture of magnetic field lines, obtained from the external
uniform field by accretion of a non-rotating gas with infinite conductivity into a black
hole. Dash lines are produced by the self-similar accretion flow with a sink in
the equatorial plane. In reality the accretion gas is forming a disk in the equatorial plane,
where matter is collected, and is moving into a black hole due to
non-perfect conductivity. This motion is generating additional azymuthal
currents in the disk, changing the external magnetic field.
Solid lines give the self-consistent magnetic field structure
with account of currents in the disk, from \cite{bkruz76}. }
\end{figure}

 Rotation of
matter in such a disk is accompanied by generation of a strong
electrical field, leading to particle acceleration and ejection of
matter along magnetic field lines. Qualitative and
phenomenological models had been considered in \cite{bkb76},
\\ \cite{blan76}, \cite{lov76}. Analytical self-similar solutions
for jets and winds produced by centrifugally and magnetically
driven mechanisms had been obtained in \cite{bp82},
\cite{lovbc91}.

Extensive numerical simulations have been done for construction of
more realistic models of quasi-stationary jet formation from
magnetized accretion disks. In the paper \cite{ruk98}
 numerical simulations have been done  of
dynamics of magnetic loops in the coronae of accretion disks. It
was obtained that in presence of differential rotation the loops
are opened in the inner parts of the disk, and opening of loops is
followed by magnetically driven outflow from the disk. The outflow
may be transient, consisting of small-scale outbursts from
different loop opening, or may be steady, corresponding to outflow
along the open field lines at the inner part of the disk, and
showing some collimation. Computations of magnetocentrifugally
driven winds had been performed in \cite{ukr99}.
The stationary regime of the outflow was
obtained by solution of time-dependent MHD equations, with a
split-monopole poloidal field configuration frozen into the disk.
Close to the disk the outflow is driven by the centrifugal
force, while at all larger distances the flow is driven by the
magnetic force, which is proportional to $-\nabla(rB_{\phi})^2$,
where $B_{\phi}$ is the toroidal field. The collimation distance
over which the flow becomes collimated is much larger than the
size of the sumulation region, so the obtained outflows are
approximately spherical.

MHD simulation have been performed in \cite{ulr2000}
of Pointing outflows, where the mass flux is
negligible and energy and angular momentum are carried
predominantly by electromagnetic field in the form of MHD waves.
As a result of time-dependent simulations a quasi-stationary
collimated Pointing jet arises from the inner part of the disk and
a steady uncollimated hydromagnetic outflow from the outer disk
for a case Keplerian disk initially threaded by a dipole like
poloidal magnetic field.

\section{Magnetorotational explosions}

 Mechanism of magnetorotational explosion
was suggested in \\ \cite{bk70}, where amplification of the
to\-ro\-idal component of the magnetic field due to differential
rotation leads to the transformation of a part of the rotational
energy to the energy of the ejection. First numerical simulations
devoted to the magnetorotational mechanism were made in
\cite{lbw}, and investigated analytically by \\ \cite{meier}.

One of the most important parameters for this problem is relation
of magnetic energy to the gravitational energy of the star:
$\alpha=\frac {E_{\rm mag}}{\vert E_{\rm grav}\vert }$.
2-D simulations have been done in \cite{ar2000} for the initial values of the
$\alpha=10^{-2},\>10^{-4},10^{-6}$. The computations have
been performed using implicit Lagrangian scheme with triangle
reconstructive grids, specially designed for astrophysical
problems.
 Details of the definition of the initial magnetic field
are described in the paper \cite{ar2000}.

Initial magnetic field chosen for our simulations has
quadrupole-like kind of symmetry (i.e. its "z"- component is equal
to zero at the equatorial plane). As a result an ejection
predominantly in the equatorial plane was obtained. The toroidal
magnetic field grows with time and produces MHD shock which moves
to the boundary of the star. Part of the matter of the envelope of
the star (about 7\% of the mass of the star)  has radial kinetic
energy larger than its potential energy and can be ejected. This
ejected matter carries about 3.3\% of the total energy of the star
(see also the previous contribution). When the initial magnetic field have a
dipole-like symmetry jet formation is possible as was first shown in
calculations of \cite{lbw}. Calculations with the initial magnetic
field structure, similar to \cite{lbw} have been performed using
the same program as in \cite{ar2000}. The results are presented in
Figures 2-4. The initial model (Fig.2, left)
was constructed as a differentially
rotating star with a quasi-dipole magnetic field (Fig.2, right), produced
by toroidal currents ring situated at about 0.5 of the equatorial
radius, where density was less than 0.1 of the central density.

\begin{figure}
\resizebox{14pc}{!}{\includegraphics{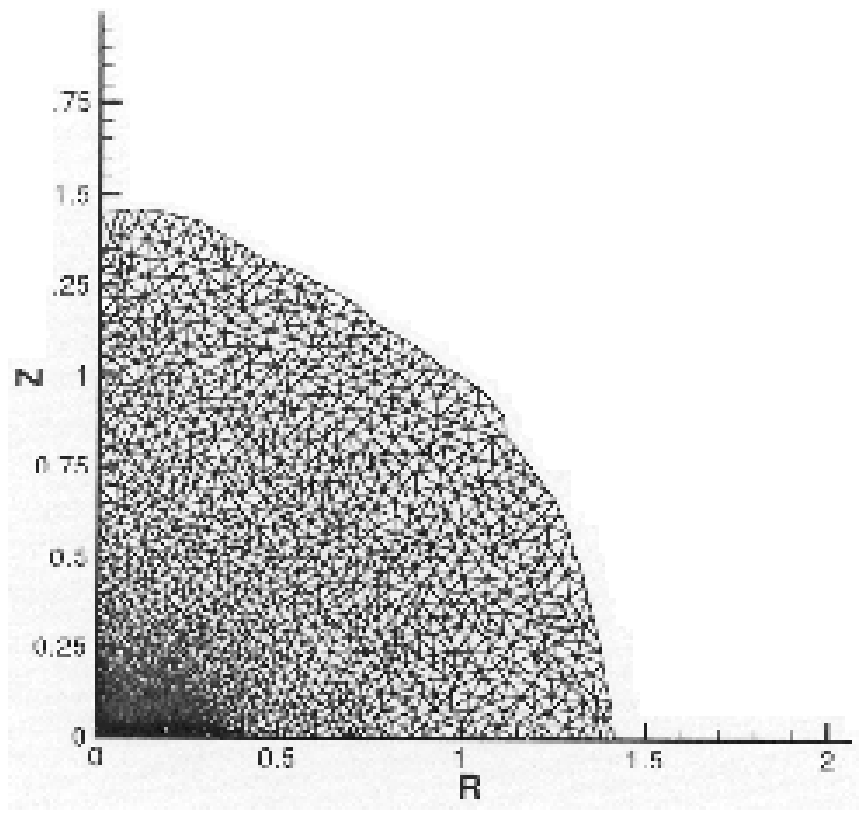}}
\resizebox{14pc}{!}{\includegraphics{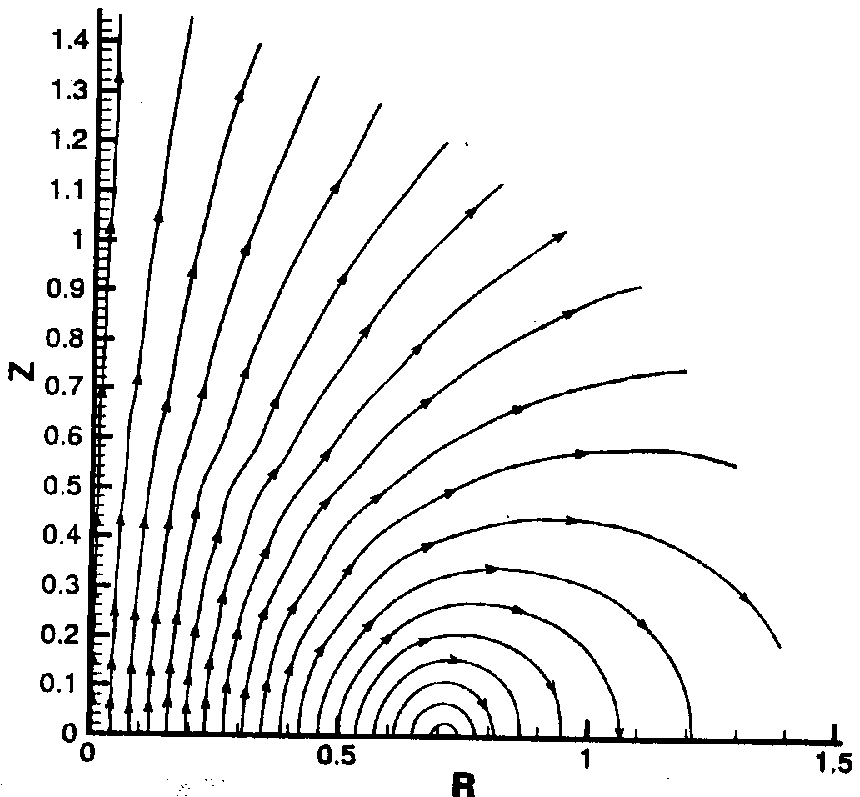}} \caption{Lagrangian
grid (left), and poloidal magnetic field distribution (right) in a
stationary equilibrium differentially rotating star at
$t_1=18.918451$, when the magnetic field was included into
calculations. }
\end{figure}

The time is given in non-dimensional units, one unit of time is roughly
equal to the time of crossing of stellar radius with the parabolic speed
of the initial model.
Initial distributions of the density and angular velocity  are
represented in Figure 3.

\begin{figure}
\resizebox{14pc}{!}{\includegraphics{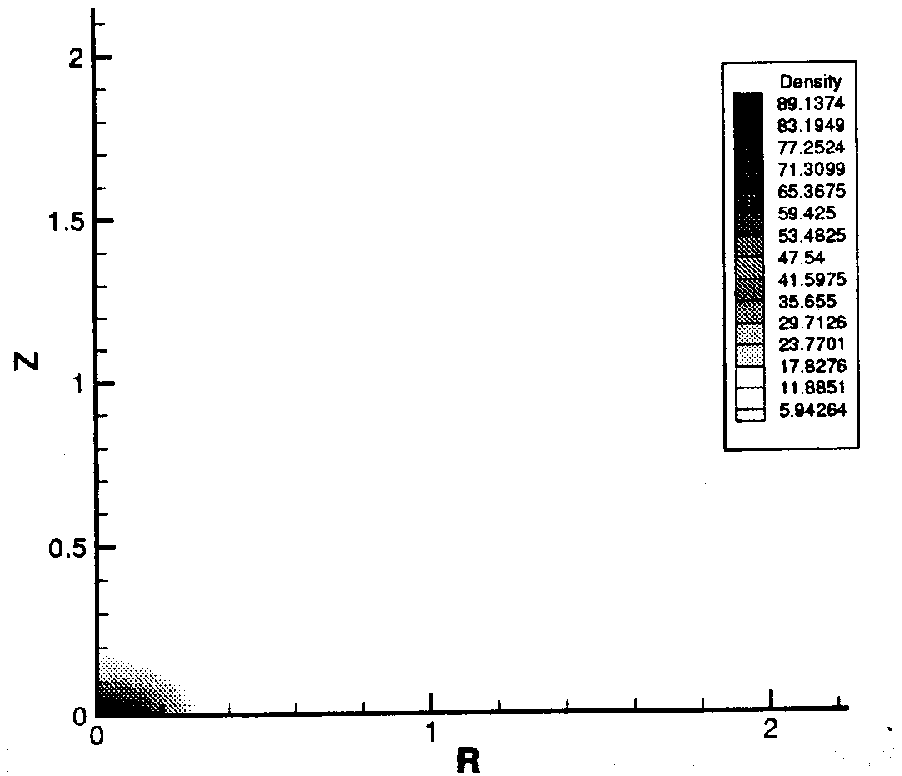}}
\resizebox{14pc}{!}{\includegraphics{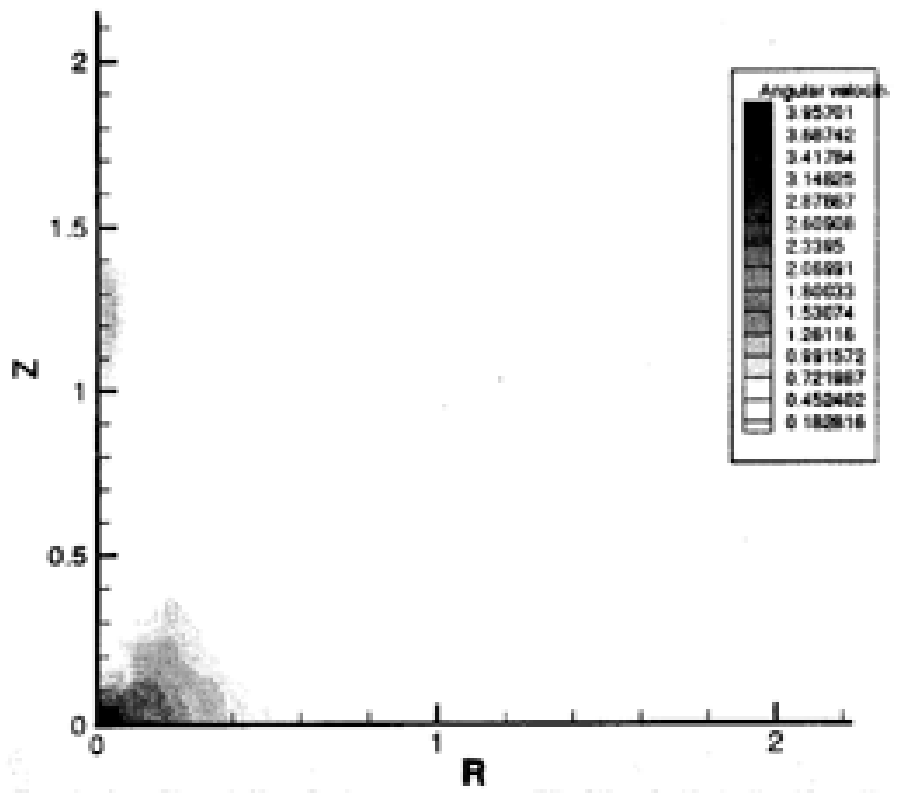}} \caption{Density
(left), and angular velocity (right) distributions at
$t_1=18.918451$.}
\end{figure}

Magnetic field twisting produced strong toroidal magnetic field
with local energy density of the order of the one of the matter.
Absence of the radial component of the magnetic field in the
equatorial plane led to formation of toroidal field rings at
higher latitudes. The excess of the magnetic pressure produced a
matter compression across the axis, and ejection of the matter in
the direction attached to the axis. Density distribution and
velocity field in the last calculational point with clear
indication of formation of the ejection are represented in
Figure~4.

\begin{figure}
\resizebox{14pc}{!}{\includegraphics{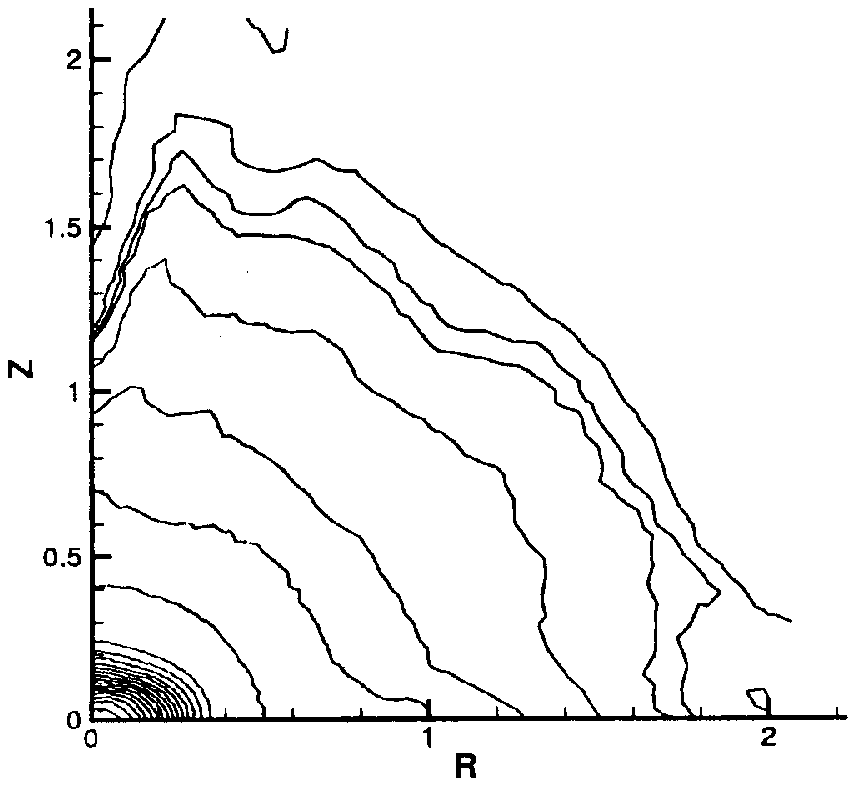}}
\resizebox{14pc}{!}{\includegraphics{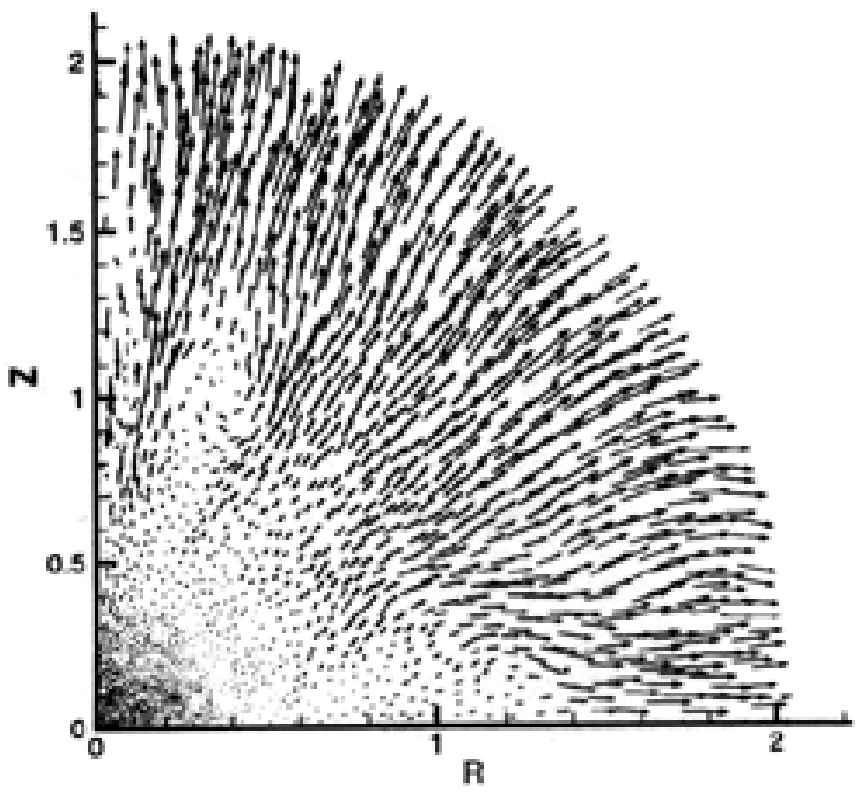}} \caption{Density
distribution (left) and  velocity field (right) at $t=30.947880$,
when the mass ejection preferentially in the pole direction is
formed.}
\end{figure}

\section{Mirror symmetry braking of magnetic field in
differentially rotating stars}

There is a possibility of asymmetric MHD explosion, when we have
asymmetric magnetic field amplification. One example of
development of asymmetric picture was considered by
\cite{bkm92}. When the col\-lapsing and
exploding star has initially toro\-idal and polo\-idal fields of
different symmetry: dipole polo\-idal and symmetric polo\-idal, or
quad\-ru\-pole polo\-idal and anti\-sym\-metric toro\-idal fields,
formation of an additional to\-ro\-idal field from the existing
po\-lo\-idal due to differential rotation leads to spontaneous
breaking of the symmetry. Due to the asymmetry of the to\-ro\-idal
field during the explosion the outbursts could also become
asymmetric, giving observational asymmetric or even one-side jets
often observed in extra\-galactic jets, and may be in
micro\-quasars \cite{mr92},
\cite{fen}. Another possibility of the asymmetry of the magnetic field
was considered in \cite{wsl92}, where disk with the mixture of
dipole and quadrupole fields was considered. Asymmetric magnetic
field in the accretion disk may lead to appearance of asymmetric
jets in stationary and non-stationary variants. In transient
accretion disks, formed by tidal capture or destruction of the
nearby star by a black hole, asymmetric magnetic field may be
formed, and magnetorotational explosion in such disk may lead to
the one side ejection.

{\bf Acknowledgments}

This work was partially supported by RFBR grant 99-02-18180 and
INTAS-ESA grant 120. The authors are grateful to Prof. J.C.Wheeler
and the Organizing Committee for support and hospitality, and to
O.D.Toropina for help.

% choose bibtex style depending on layout style and options used in
% sample:

%\doingARLO[\bibliographystyle{aipproc}]
%          {\ifthenelse{\equal{\AIPcitestyleselect}{num}}
%             {\bibliographystyle{arlonum}}
%             {\bibliographystyle{arlobib}}
%          }
%%\bibliography{sample}
%\bibliography{bkjet}
%

\end{document}